**Graphical Table of Contents for "Superlubricity of epitaxial monolayer WS$_2$ on graphene" by Büch *et al.*_**

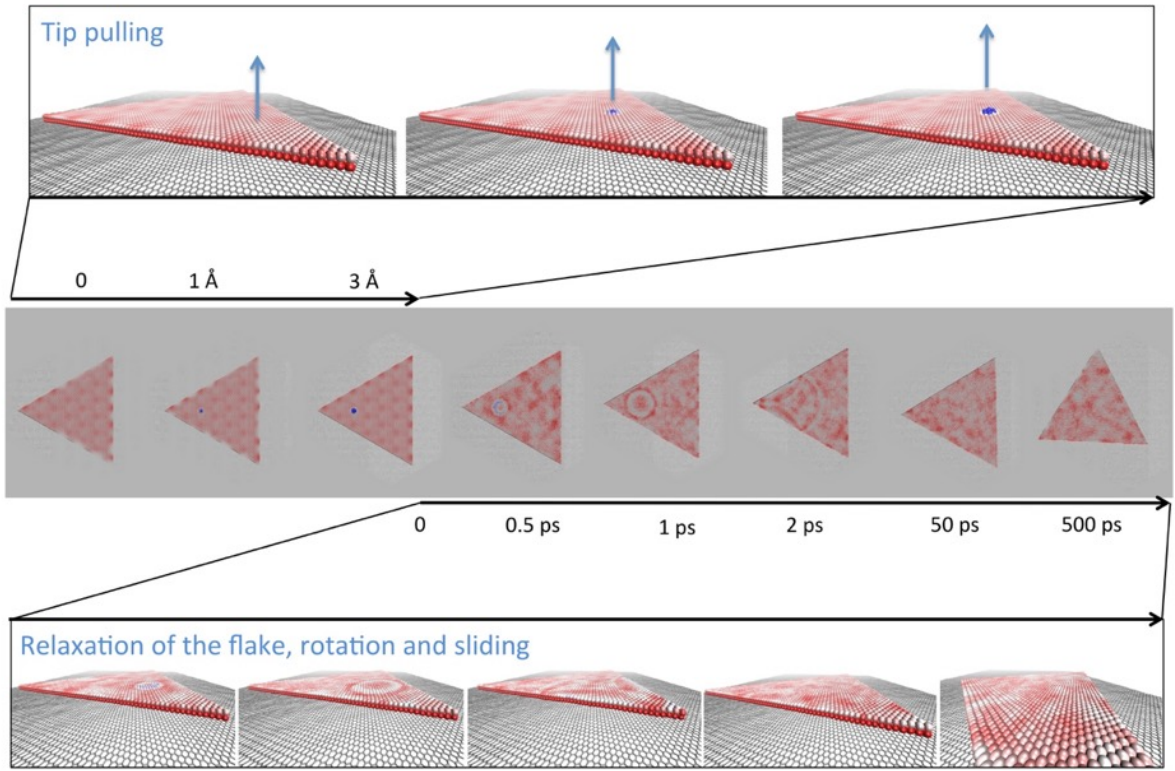

This work is a combined experimental and theoretical study reporting superlubricity of monolayer tungsten disulfide on graphene triggered by scanning probe tips.

# Superlubricity of epitaxial monolayer WS$_2$ on graphene

Holger Büch[1], Antonio Rossi[1,2], Stiven Forti[1], Domenica Convertino[1,2], Valentina Tozzini[2] and Camilla Coletti[1,3,*]

1. Center for Nanotechnology Innovation @NEST, Istituto Italiano di Tecnologia, Piazza S. Silvestro 12, 56127 Pisa, Italy

2. NEST, Istituto Nanoscienze – CNR and Scuola Normale Superiore, Piazza San Silvestro 12, 56127 Pisa, Italy

3. Graphene Labs, Istituto Italiano di Tecnologia, Via Morego 30, 16163 Genova, Italy

* camilla.coletti@iit.it

**Abstract**

We report on the superlubric sliding of monolayer tungsten disulfide (WS$_2$) on epitaxial graphene (EG) on silicon carbide (SiC). WS$_2$ single-crystalline flakes with lateral size of hundreds of nanometers are obtained via chemical vapor deposition (CVD) on EG and microscopic and diffraction analyses indicate that the WS$_2$/EG stack is predominantly aligned with zero azimuthal rotation. Our experimental findings show that the WS$_2$ flakes are prone to slide over graphene surfaces at room temperature when perturbed by a scanning probe microscopy (SPM) tip. Atomistic force field based molecular dynamics simulations indicate that through local physical deformation of the WS$_2$ flake, the scanning tip releases enough energy to the flake to overcome the motion activation barrier and to trigger an ultra-low friction roto-translational displacement, that is superlubric. Experimental observations indicate that after the sliding, the WS$_2$ flakes rest with a rotation of n$\pi$/3 with respect to graphene. Atomically resolved investigations show that the interface is atomically sharp and that the WS$_2$ lattice is strain-free. These results help to shed light on nanotribological phenomena in van der Waals (vdW) heterostacks and suggest that the applicative potential of the WS$_2$/graphene heterostructure can be extended by novel mechanical prospects.

1. Introduction

Thanks to their enticing electronic, optical and magnetic properties, two-dimensional (2D) van der Waals (vdW) heterostacks are a playground for fundamental studies and display a wide range of potential applications [1,2]. Among the broad portfolio of possible 2D heterostacks, those obtained by combining transition metal dichalcogenides (TMDs) and graphene are presently under the spotlight owing to their appealing optoelectronic properties. In particular, the TMD tungsten disulfide (WS$_2$) exhibits large carrier mobility [3], high photoluminesce emission [4], significantly large spin-orbit splitting (~ 460meV) [5], and long exciton life and coherence times[6]. Novel field-effect vertical tunneling transistors [7] and photodectors [8,9] based on WS$_2$/graphene stacks have been demonstrated. While the optoelectronic potential of this 2D heterostack is being extensively studied, minor attention has been paid to its nanotribological properties. Both graphite and bulk WS$_2$ are known as effective lamellar lubricants with hexagonal structures. During the years, a number of tribological experiments have been carried out on both graphite and bulk molybdenum disulfide (MoS$_2$), whose structure is analogous to WS$_2$. Interestingly, novel phenomena as self-retraction and superlubricity have been reported [10–13].

The term superlubricity, indicating vanishing friction, was first introduced in the early 1990s by Hirano et al. [14–16] who related friction to absence of "commensuration" between crystal surfaces. This was in turn defined in terms of the difference between lattice parameters, symmetry and relative rotation of surfaces [14]. Parameters related to commensuration were later defined, such as the "registry index" [17], measuring the rigid geometrical superposition of atoms at the (possibly rotated) surfaces. However, it was later recognized [18], that the relaxation of the atoms at the surface has a fundamental role, that is, it introduces an energy dissipation mechanism. The dissipation was related to surfaces flexibility [18] by the parameter $\lambda=u/(Ka^2)$, measuring the relative strength of the inter-surface interaction $u$ (usually the vdW interaction) to the rigidity of the surfaces $K$, with $a$ being the lattice parameter. A system would be prone to superlubricity if the surfaces in contact are rather rigid (large $K$) and/or the inter-surface interaction $u$ is small. Indeed, the effect of vertical flexibility and relaxation of the surfaces was shown to reduce the role of commensurability in the lubricity phenomenon [19], by conversely giving a role to flexural phonons. The phenomenon of superlubricity displays several non-trivial aspects, and no broad consensus has been reached yet, even on the definition of the term itself. In this work, the term superlubricity is exploited to describe a 2D flake sliding over a substrate without the need for a horizontal force and in a non-diffusive regime.

In this sense, superlubricity, was observed in a graphene flake attached to an atomic force microscopy (AFM) tip sliding over a graphite substrate [20]. It was also reported in scanning tunneling microscopy (STM) experiments at temperatures as low as 5 K [21]. Very recently, friction tests on the sliding of a single-layer MoS$_2$ nanoflake on another incommensurate misaligned MoS$_2$ substrate have yielded ultra-low friction coefficients of about 10$^{-4}$, indicating superlubricity [22]. Similarly, extremely low friction was measured when sliding a multilayer graphene–coated microsphere on graphite and bulk hexagonal boron nitride (hBN) [23]. Also, rotation and translation of graphene flakes on hBN via AFM-tip interaction has been previously reported [24]. A number of theoretical studies have predicted superlubricity in vertical homo- and heterostacks of 2D materials [25,26]. To date, however, superlubricity between a TMD and graphene – a natural playground for nanotribological studies – has not yet been experimentally observed.

Here we report the superlubric sliding of monolayer WS$_2$ nanoflakes over single-layer epitaxial graphene (EG) triggered by an STM tip. WS$_2$ was synthesized on EG on silicon carbide (SiC) via chemical vapor deposition (CVD) [27]. In comparison to heterostacks obtained via mechanical exfoliation, this entirely bottom-up heterostack presents a WS$_2$/graphene interface free of resist and solution residuals, and it is thus ideal for nanotribology investigations. We interpret the reported experimental results by means of classical molecular dynamics simulations which indicate that the simple interaction of a WS$_2$ flake with a scanning tip is sufficient to trigger its sliding.

## 2. Experimental

Graphene was obtained via solid state thermal decomposition of atomically flat nominally on-axis 6H–SiC(0001) using a resistively heated cold-wall reactor [28]. The as-grown graphene/SiC samples were then used as receiving substrates in a tube-furnace low pressure CVD reactor, while sulfur and tungsten trioxide powders served as chemical precursors and argon as carrier gas [27]. STM measurements were carried out in an Omicron low-temperature STM system. After transfer to the ultra-high vacuum (UHV) chamber, the sample was annealed via resistive heating to 400 °C for about 12 hours. The STM was operated in constant current mode using an etched tungsten tip. The tip voltage was varied between 0.1 V and 2.7 V, and the tunnel current between 10 pA and 300 pA. The pressure in the STM chamber was better than 5 x 10$^{-11}$ mbar. Scanning electron microscopy (SEM) imaging was performed at 5 keV using a Zeiss Merlin microscope, equipped with a field emission gun. Atomic force microscopy was performed using an AFM+ by *Anasys*. The Gwyddion software package was used for the analysis of the STM and AFM images [29]. The AFM videos in the Supporting Information (SI) are built with subsequent AFM scans. Low energy electron diffraction (LEED) measurements were carried with an ErLEED by SPECS. All measurements were performed at room temperature (RT).

Molecular dynamics simulations were performed with a model system including a rhombohedral graphene sheet of ~32 nm side and WS$_2$ triangular flakes of ~5nm, ~10 or ~20 nm side. In order to speed up the simulations, a minimalist connective model, the elastic network, was used to represent the internal dynamics of the graphene sheet and the flake. The interaction parameters were optimized in order to reproduce experimental data or data from higher order theories. The flake-substrate interactions were represented via Lennard Jones (LJ) potential with standard parameterization. A preliminary evaluation of vdW energy performed by means of a geometry optimization on the flat substrate indicated a vdW energy per atom of about 110 meV/atom, of the same order but larger than the value obtained in our previous work by means of Density Functional Theory calculations [5]. This discrepancy might be due to different causes (registry and corrugation, underestimation of vdW energy in DFT), therefore, in absence of a direct evaluation of the experimental value of this parameter, we use the standard values of the vdW parameters, widely used and tested. The whole parameter set, as well as more details about the model systems and the force field, are reported in the SI. Simulations were performed with the code DL_POLY [30], using in-house programmed software tools to create the input files and analyze the output. The simulation timestep was set to 1 fs and standard algorithms were used for integration of equations of motion (Verlet, and Nosé thermostat for constant temperature runs). Simulations were run on an 8-core workstation.

## 3. Results and discussion

### 3.1 WS$_2$ sliding triggered by an SPM tip

In Fig. 1(a), we show a large-scale filled-state STM image of as-grown triangular WS$_2$ flakes on the graphene/SiC substrate. Within the field of view, a few WS$_2$ flakes are visible with an apparent height of ~7 Å (see line profile in the inset), confirming the monolayer thickness of the TMD. Most of the flakes lie smoothly across the SiC step edges (~8 Å in height). The size distribution of the WS$_2$ flakes is appreciably narrow. The average side length of the flakes analyzed in the sample is about 600 nm, corresponding to an area of about 0.16 µm$^2$, as confirmed by scanning electron microscopy (SEM) analysis (see the histogram reported in Fig. 1(d)). Larger areas reported in the histogram are typically caused by merging of triangular grains (i.e., detected values are multiples of those retrieved for isolated crystals). A simple inspection of the STM and SEM micrographs indicates the existence of an epitaxial relation to the substrate. Indeed, most of the triangular flakes present parallel edges. As evidenced in panel (c), only a limited number of flakes – i.e., about 10% – were found to present edges rotated by 30°. LEED analysis – performed on a macroscopic area of about 1 mm$^2$ – is shown in Fig. 1(b) and reveals that the majority of the WS$_2$ flakes are perfectly aligned with the substrate as the highest intensity WS$_2$ diffraction spots are 0° oriented with respect to the graphene diffraction spots. We also note a faint complete circle in the LEED image belonging to WS$_2$ reciprocal lattice vectors, which we attribute to a low percentage of WS$_2$ nanoflakes of stray orientation. Our result differ from what reported from Miwa et al. [31] for MoS$_2$/graphene heterostacks synthesized in vacuum, where the TMD was found to be rotated 30° with respect to the graphene BZ. On the other side, a similar flake orientation distribution was already observed for CVD-grown TMD/graphene [5,32], which suggest that the vapor-phase approach is a suitable technique to obtain heterostructures with azimuthal alignment.

In Figs. 2(a)-(c), we show a typical behavior observed when scanning with an STM tip over sample areas presenting WS$_2$ flakes. The three STM images of the same surface area display the same WS$_2$ flake in three different positions (marked by arrows) at subsequent scans. Initially, the WS$_2$ flake is located in the bottom right side of the image (panel (a)). In the subsequent image the flake has moved to the central left area of the micrograph (panel (b)) and finally, in the third scan, is found in the bottom left side (panel (c)). Hence, it appears that upon triggering of the STM tip the flake slides by rotating and translating in random directions, not connected with the scanning direction. It should be mentioned that the sliding of the WS$_2$ flake on top of the graphene substrate is not always observed when performing STM imaging. The scanned WS$_2$ flake might either sit still or move suddenly to another position, often several micrometers away (i.e., outside the field of view of the scan) with what appears to be a stochastic trigger, i.e. flake sliding becomes more likely the longer we scan over it. Fig. S1 reports what typically observed for larger scan areas, where some of the flakes sit still while others slide around without preferential direction. The locked flakes are often part of large clusters of flakes or sit across step edges. When sliding, after each transition, the WS$_2$ flakes are always found to sit in commensurate position with respect to the underlying graphene layer. They either maintain the same initial orientation or rotate by 60°, reasonably indicating the most stable stacking.

We should point out that our observation of a sudden and fast sliding of a WS$_2$ flake over a graphene substrate is in disagreement with the experimental findings reported for similar systems, namely MoS$_2$ on epitaxial graphene and MoS$_2$ on graphite [32-34]. However, WS$_2$ on epitaxial graphene has already been reported to differ on several aspects, all of which point towards the indication of the WS$_2$ as a less interacting material with epitaxial graphene on SiC, than MoS$_2$ [35,5].

In our experiments, we do not observe a trend in the moving probability as a function of tip voltage (while keeping the tip-sample distance constant). This rules out electrostatic effects being the triggering cause of the sliding as it is instead reported in electrostatic-manipulation STM (EM-STM) experiments [36,37]. A further confirmation of this is given by the observation of WS$_2$ non-directional sliding triggered by an AFM probe. When scanning in contact mode, some of the WS$_2$ flakes rotate and translate as shown in Fig. S2, videos S6 and S7 and in agreement with the STM measurements. In contrast, no WS$_2$ flake movement is observed in tapping mode, which we attribute to the strongly reduced time during which the vdW forces between tip and sample reach the full strength. Since during the AFM scan both sample and probe are grounded, electrostatic manipulation is not at the basis of the observed sliding. Also, these observations are significantly different from those reported by Kobayashi et al., where WS$_2$ grains are manipulated and physically pushed with a nano-prober system [38].

A detailed analysis of STM and AFM subsequent scans indicates that sliding flakes typically come to a stop at the edge of steps (or when approaching flake clusters), until they are triggered again by the tip (see SI). Hence, typically sliding distances for the WS$_2$ flakes range between 500 nm to a few micrometers (i.e., the terrace widths for the samples analyzed). Also, we note that flakes can both climb up and climb down steps. Statistically we find that climbing up is more likely to happen on step with lower heights (i.e., bilayer slabs of SiC rather than unit-cell-high steps – see SI).

The sliding of WS$_2$ flakes does not induce defects in the underlying graphene, as shown in panels (d)-(g). Panel (d) shows a larger scan to clearly display the initial presence of the WS$_2$ flake, and panel (e) is the zoom-in to one of its edges. In the micrograph reported in panel (g), acquired immediately after that reported in panel (e), the flake appears to have slid away. The corresponding zoom-in (panel (f)) shows that the graphene layer where the WS$_2$ was previously located retains its lattice with no alteration of the atomic registry. The sliding behavior observed in this work appears very similar to what was reported in reference [21] where superlubric sliding of graphene flakes on graphite was triggered by an STM tip. To further clarify the triggering mechanism for the specific system molecular dynamics simulations – whose main results we now discuss – were carried out.

### 3.2   Molecular dynamics simulations modeling

Before investigating the tip-triggered motion, we perform simulations aimed at investigating whether the motion of WS$_2$ flakes on graphene can be thermally activated. We performed simulations at different temperatures for different flake sizes (see the SI for a detailed list of simulations). For all flakes sizes, we observe an oscillating horizontal motion of the flake at low temperatures, while translational motion is activated at higher temperature, in some cases preceded by an oscillating-percolation phase, in which we also observe two main preferential orientations, namely 0° (and its equivalents) or 30° (and its equivalents) (see Fig. 3(a),(c) for a flake of ~10 nm side at ~200 K and plots and movies in the SI). The translational motion activation temperature $T_c$ depends on the size: we observe a clear trend of dependence on the squared root of the number of atoms (or flake area), or equivalently, linear dependence on the side of the flake (see plotted data points in Fig. 3(b)). Therefore, translational motion is thermally activated for flakes with lateral size of a few tens of nanometers even at room temperature, while larger flakes appear steady. Interestingly, the oscillatory phase duration depends on temperature and flake size.

This result can be explained within the framework of the general theory phase transition. In general, a transition occurs when $k_B T$ overcomes a given energy barrier $\Delta E$. This value, for the thermal activation, is related to the *static* friction coefficient, and roughly corresponds to the flake – substrate interaction per unit surface. A rough estimate can be obtained from the average vdW energy per contact atom $\varepsilon$: simulations return -108 – -112 meV, also depending on the temperature (see Fig. 3(c)), pointing to activation temperatures exceeding 1000 K. However, this simplified picture assumes the thermodynamic limit, i.e. very large flakes. For small flakes the transition is driven by the amplitude and correlation length of energy fluctuations (diverging near $T_c$) rather than by the average value of the energy. Because fluctuations scales as the squared root of the number of particles, this explains the observed behavior. The transition temperature as a function of the size fitted according to this rationale on simulation data is reported in Fig. 3 (b). Indeed, according to our simulation data, flake with lateral dimension of 20-50 nm are expected to be mobile at RT. Instead, flakes of 100 nm side have thermal activation temperature above RT. The thermal activation at RT of nanometer-sized flake might well explain the lack of visualization of such flake population in our experimental analysis.

Hence, for the flakes experimentally studied in this work, which present a lateral size of about 600 nm, we can exclude thermal fluctuation as the cause for motion. The second set of simulations aimed at demonstrating that, conversely, their motion can be activated by a pulse delivered by the tip. We describe the event as a pulling of the tip of an area of the flake up to a given height with consequent deformation of the flake. Specifically, a flake with a lateral size of 20 nm was pulled away from the substrate by a distance of 3 Å (see Fig. 4, top panel). After being locally released from lifting, as it physically happens while the tip scans over the specimen, the

"sliding barrier" is overcome and the flake slides over the graphene surface (Fig. 4, middle and bottom panel). Indeed, when the tip detaches, the deformation energy is released, and transferred in translational motion after some flexural like oscillation propagating from the tip pulling location. The transition phase lasts for ~200 ps, after which the flake starts to rotate and then to translate, at the starting velocity of 7 pm/ps. The flake velocity upon motion activation is found to be 2-4 pm/ps. The total transferred energy from the tip to the flake is more than ΔE~2 eV, which is subsequently distributed over the flake atoms, allowing to overcome the motion activation barrier in less than a few nanoseconds times. We observe that the transferred energy ΔE depends on the deformation of the flake due to the pulling, which is basically independent on the size of the flake, being local. Therefore, it is expected that smaller flakes are activated more easily and faster from the tip, while larger flakes might take larger times or need stronger pulling to be activated in microscopic times. In fact, more efficient momentum transfer from the tip might occur in dragging of pushing type interactions (in simulations only a pulling type event is considered).

We would like to point out that once the static friction barrier is overcome, the flake undergoes in a different regime where the dynamic friction comes into play. This was estimated by means of a set of simulations where the flakes are forced to slide horizontally by applying a constant force per unit surface, and overdamped, so that they adiabatically follow the energy profile (Fig. 3(d) and the SI). In this case fluctuations are small, and the energy profile is substantially independent on the size. As it can be seen from Fig. 3(d) the profile follows the super-periodicity of the moiré pattern with local fluctuations due to the atomic structure, and the barriers are of the order of 0.5-1 meV, much smaller than the motion activation barrier and than $k_BT$. This means that once activated, motion proceeds with ultra-low friction at RT, in a superlubric fashion.

However, it has to be noted that in the realistic case of a surface with atomic steps, the superlubric motion is hampered by the presence of such atomic corrugations. As experimentally, observed, step edges are the preferential (often temporarily) locking position for flakes. Indeed, as it can be visualized in the SI AFM videos in some instances the $WS_2$ flakes overcome atomic steps and continue their sliding. In order to explain this, it is necessary to point out that on SiC, graphene grows in a carpet-like fashion [39, 40] and therefore the step the $WS_2$ flake faces is less abrupt, i.e. "atomically smoothened" by the presence of a continuous graphene overlayer. Hence, the step is seen from the flake as a rather sweet foot of a hill, with a local curvature, which is not particularly strong (i.e., the flake doesn't have to suddenly climb 7.5 Å). From the microscopic point of view, overcoming corrugation is favored by the fluctuations, which bend the edges and vertices of the flake. Simulations show that, at RT, the root mean squared fluctuations of atoms in the center of the flake are less than 1 Å, but can increase up to 1-2 Å at the vertices (or 3-5 Å when the motion is triggered by the tip). Generally, the dissipation occurs by releasing an amount of momentum at a corrugation, which depends on is amplitude: it is very small in the quasi-flat case, leading to high lubricity, while it may increase for large level corrugation, up to a complete stop in given cases.

## 3.3 Atomically resolved imaging of WS$_2$ on graphene

The reported sliding of WS$_2$ on graphene makes atomic resolution imaging of the heterostack extremely challenging. It should be mentioned that to date, no STM study has yet investigated the WS$_2$/graphene heterostructure, while WS$_2$ nanoclusters have been studied on Au(111) surfaces [6,41]. STM images for the more complex Mo$_{1-x}$W$_x$S$_2$ on graphite were reported in reference [42]. We report that in order to obtain atomically resolved micrographs, it is crucial to image small-scale frames entirely located within a WS$_2$ flake and not including a flake edge. In general, we observed an increased sliding probability when the tip was positioned for an extended period of time in close vicinity of an edge of the flake. In simulations we observe that fluctuations are always larger at the edges rather than in the center of the flake, therefore it is reasonable that the effect of the tip is stronger if applied laterally. In Fig. 5(a) we report an atomically resolved STM micrograph obtained with a sample bias of 0.1 V and a tunneling current of 0.25 nA. A number of defective sites can be observed, which can be reasonably attributed to sulfur vacancies, generated either during the CVD growth process or to exposure of the sample to the electron beam for SEM imaging [43–45].

The epitaxial alignment of the two crystals, which exhibit different lattice parameters (i.e., 3.15 Å for WS$_2$ and 2.46 Å for graphene [5,46]), could be prone to let a moiré pattern emerge. As a matter of fact, a number of studies investigating similar systems, i.e. MoS$_2$ on graphene or graphite, evidenced the presence of a moiré as well as modifications in the TMD electronic band structure [34,35,47]. However, in our case, no moiré pattern is observed as demonstrated by Fig. 5(a). According to our molecular dynamics simulations, a moiré superlattice emerges essentially only at 0 K. At higher temperature, the superpotential is disrupted by fluctuations and at 300 K no longer visible. This lack of moiré visualization is in agreement with what reported in reference [5], where no evidence of moiré superlattice features was observed in the band structure of the WS$_2$/graphene heterostructure, further confirming the weak interaction between WS$_2$ and graphene. The inset in panel (a) report a 2D Fast Fourier Transform (FFT) filtered zoomed-in image. We assign the brightest spots to the S atoms of the top-layer of the S-W-S sandwich structure and the lower intensity ones to the W atoms. Intuitively, this assignment would be in contradiction with the fact that in WS$_2$, the density of state contribution of the metal atom is significantly stronger than that of the S atom. However, simulated STM images for the filled states [48] indicate that geometrical effects dominate in such TMDs (the metal atoms are positioned about 1.5 Å below the S atoms). Hence, a stronger tunnel current contribution by the S atoms is obtained, which leads to the observed triangular pattern. By performing an FFT of the STM image, we obtain the hexagonal Brillouin zone of the WS$_2$ structure with a lattice parameter of $3.2 \pm 0.1$ Å, as shown in Fig. 5(b). This value is in agreement with what reported for strain-free WS$_2$ [49], further indicating that the WS$_2$/graphene is a low-interacting system.

## 4. Conclusions

In summary, we have reported on the superlubric sliding of monolayer $WS_2$ on epitaxial graphene triggered by scanning probes. The studied $WS_2$/graphene heterostack presents a predominantly 0° azimuthal alignment and an atomically sharp interface as demonstrated by microscopic and diffraction studies. Sliding of the $WS_2$ flakes is observed with a high incidence although it appears to be stochastically triggered. Molecular dynamics calculations indicate that motion is triggered by the tip-sample interaction and confirm that $WS_2$ flakes slide on graphene with ultra-low friction. Thermal activation is found to be a triggering cause of motion only for flakes with lateral size one order of magnitude below those experimentally studied. The observation of superlubric sliding of $WS_2$ flakes on graphene is a further indication of the cleanliness of the CVD growth of $WS_2$, since contaminants like hydrogen can hinder superlubricity [50]. AFM analyses reveal that superlubric sliding is obtained also in ambient conditions. These findings suggest novel applicative prospects for the $WS_2$/graphene heterostack in the field of nanomotors, nanoelectromechanical systems (NEMS) and advanced optoelectronic multifunctional devices.

**Acknowledgements**

We wish to thank Professor Annalisa Fasolino for useful discussions and suggestions. The research leading to these results has received funding from the European Union's Horizon 2020 research and innovation program under grant agreement No. 696656 – GrapheneCore1.

**Electronic Supplementary Material**

Supplementary material (movies of simulation runs at different average temperatures and of AFM scans) is available in the online version of this article at http://dx.doi.org/10.1007/*********************).

# Figures

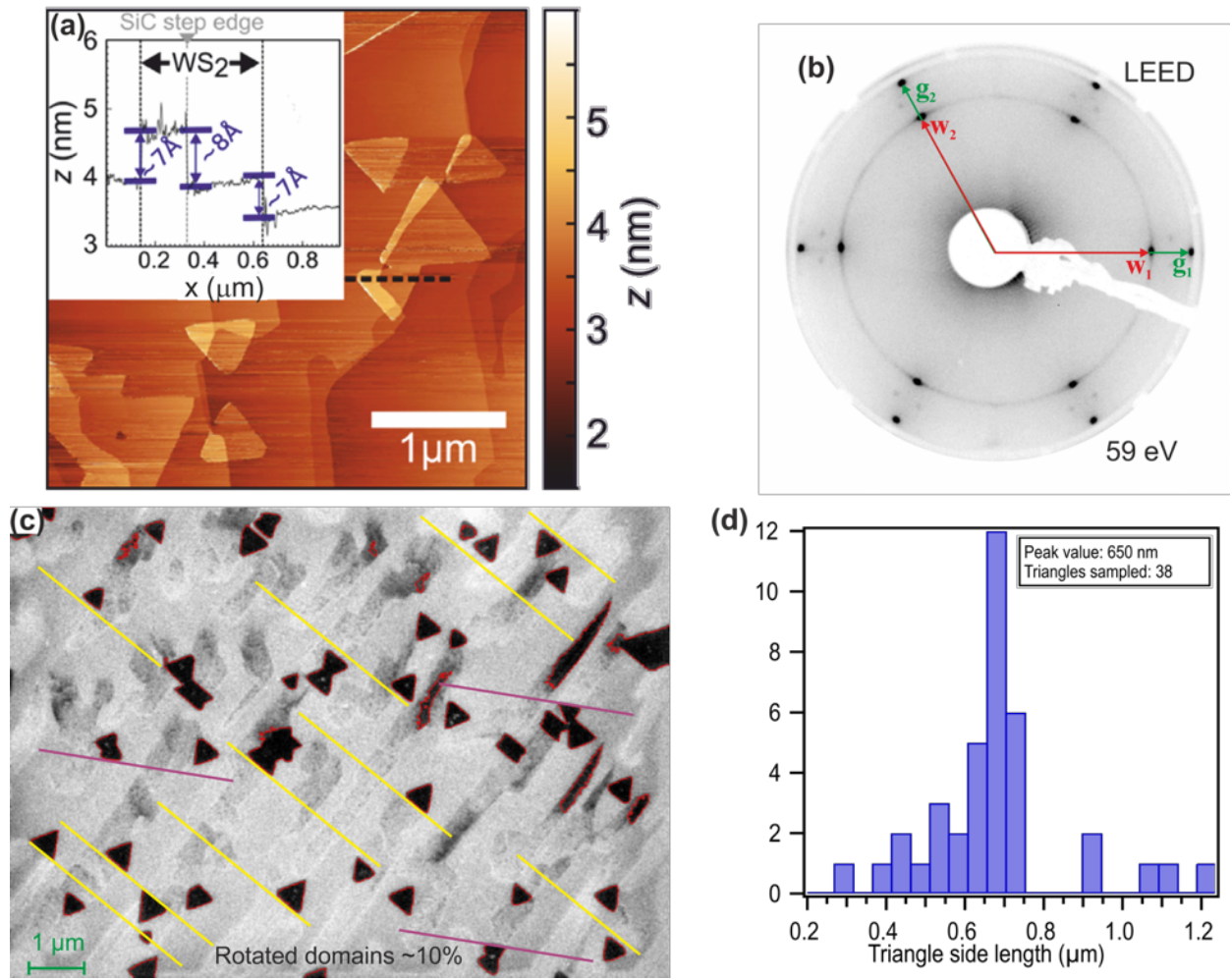

**Figure 1.** (a) Large scale STM image of WS$_2$ flakes on the graphene/SiC(0001) substrate, at a tip voltage 2.7 V and a constant current of 0.3 nA. The black dashed line marks the position of the line profile displayed in the inset. (b) LEED pattern of the WS$_2$/graphene/SiC heterostructure recorded at a primary beam energy of 72 eV. The arrows mark the different diffraction spots. (c) representative SEM micrograph highlighting the major orientations of the WS$_2$ crystals. (d) histogram of the grain side length obtained from the analysis of the SEM image in panel (c).

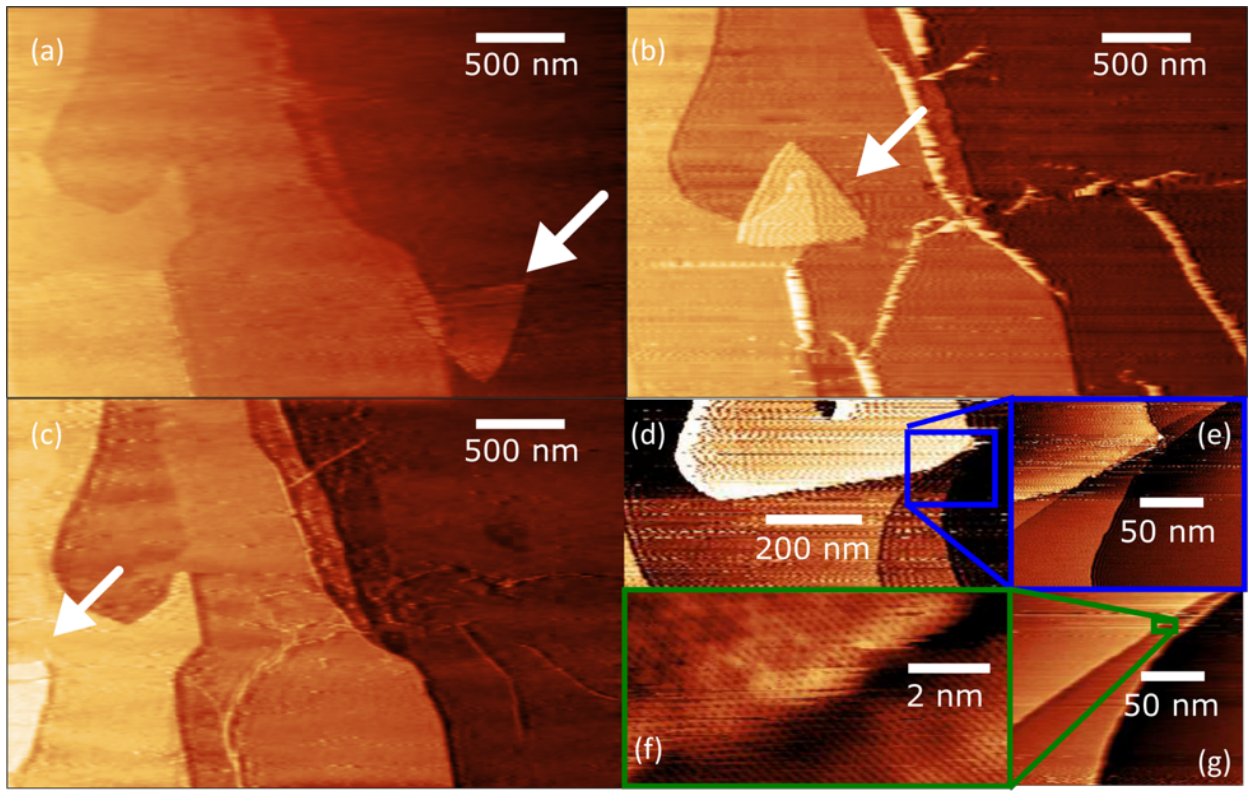

**Figure 2.** (a)-(c), Three STM images of the same area revealing three different positions of a WS$_2$ flake (marked by arrows) at subsequent scans. Tip parameters: voltage of +2.7 V and constant tunneling current of 0.3 nA. (d) STM images (2.5 V, 0.1 nA) of a surface area containing a WS$_2$ flake (partially seen in brighter color at the top end). (e) Zoom-in of the area marked by a box in panel (d). (g) Subsequent STM scan showing that the WS$_2$ flake is gone. (f) Zoom-in of area marked in panel (g) which reveals intact graphene at the former position of the WS$_2$ flake.

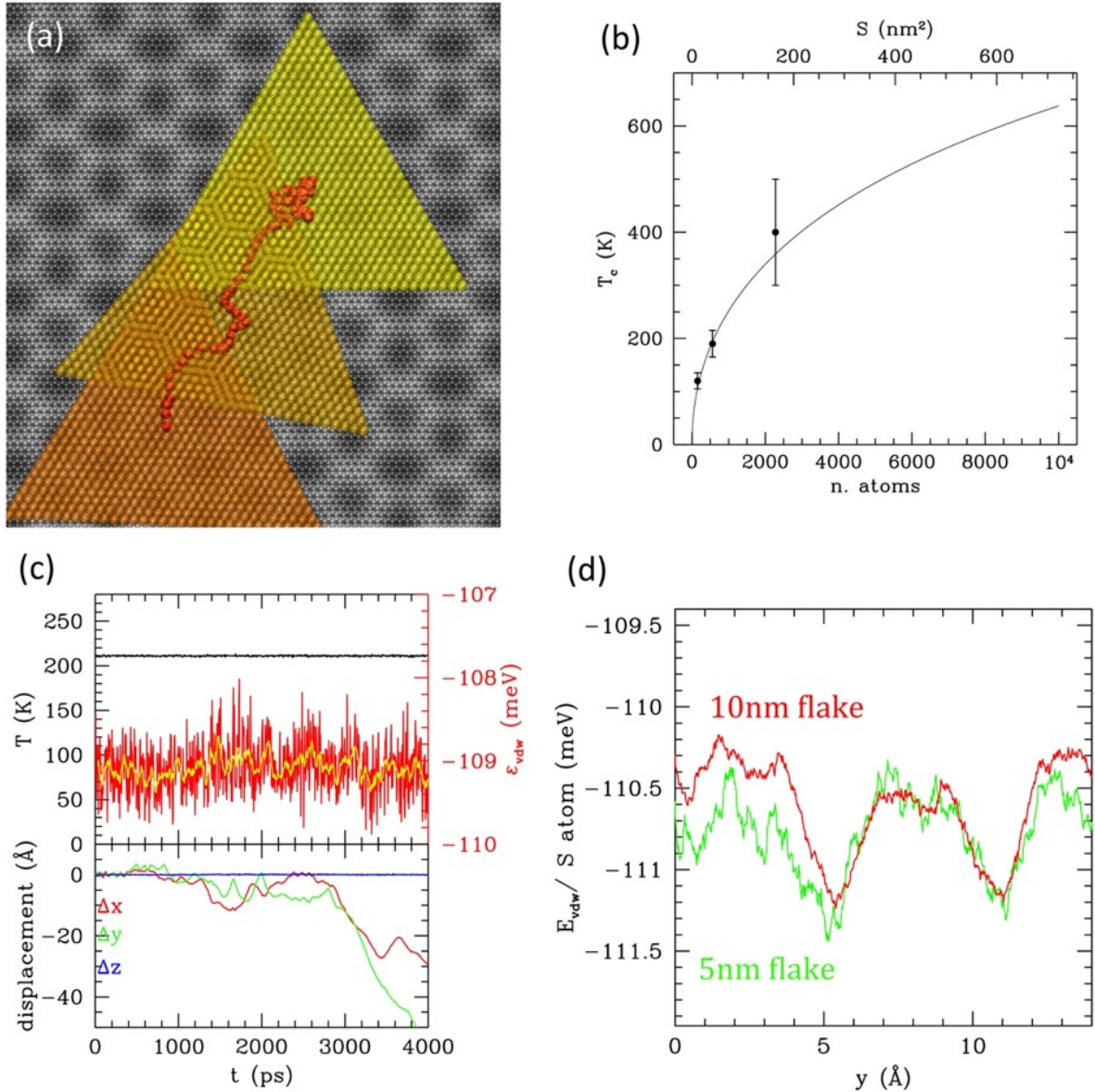

**Figure 3.** (a) Thermal activation of motion of a flake of ~10 nm at ~200 K on graphene (the corrugation of the substrate is highlighted). Red dots represent the trajectory of the center of the flake, which also undergoes a rotatory motion, as shown by a few configurations reported (yellow=first, dark orange = last). (b) Transition temperatures as a function of the flake size, as deduced by simulations. The line is a fit according to the interpolation relationship $T_c=T_\infty/(1+\surd(S_0/S))$ being $T_\infty$ the transition temperature in the thermodynamic limit and $S_0$ a parameter with the physical meaning of critical sizes for the crossover to thermodynamic limit; the number of mobile flakes with respect to the total number of flakes at a given temperature and size will then be $\rho=\exp(-T/T_c)$. (c) Temperature, vdW energy and displacement (in the three Cartesian directions) in the simulation of (a). The oscillatory phase is clearly visible up to 3ns, with a preliminary phase with very small oscillations up to 0.5ns, subsequently increasing in amplitude up to the translational motion start. The displacement in z direction is obviously always near to 0. (d) Energy profile as flakes of 5 and 10 nm slide subject to a constant force in y direction (harmchair) and overdamping (see the SI for details).

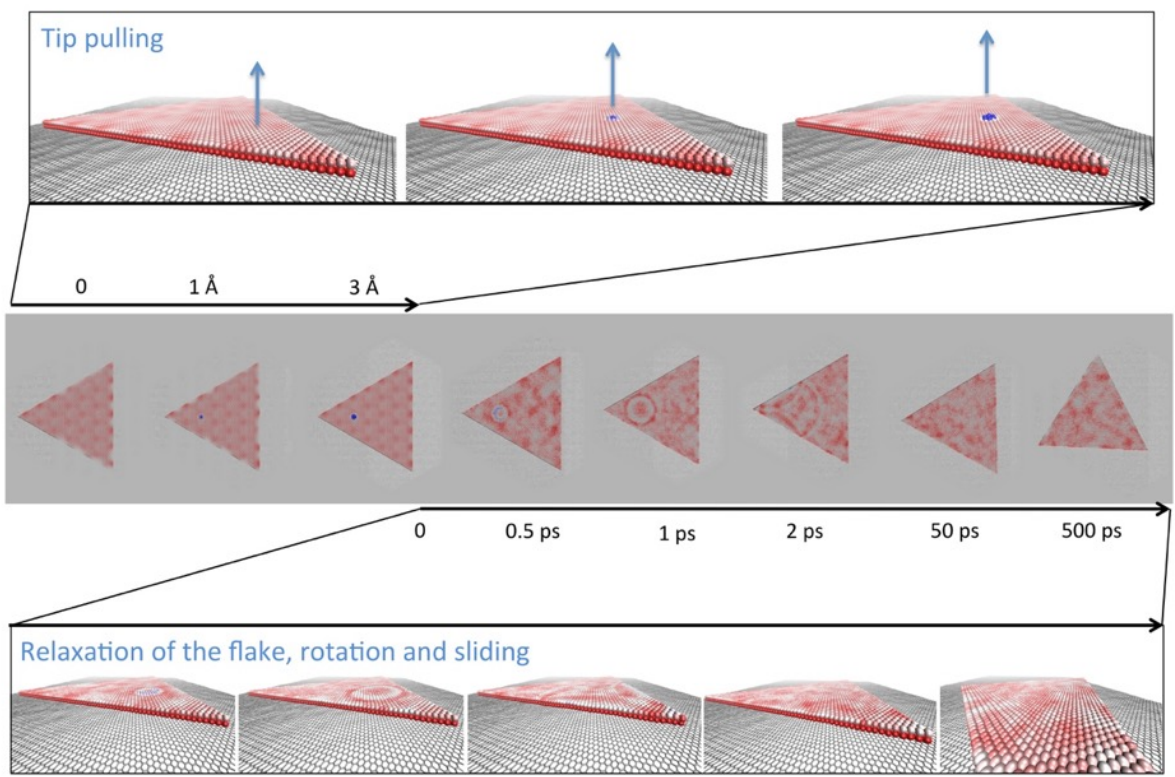

**Figure 4.** Simulation of an event of pulling of the flake with the tip up to a deformation of about 3 Å, subsequent release and free dynamics. The whole simulation lasts for 1 ns, at room temperature. Only selected relevant snapshots are reported, in top view in the central strip, in perspective in upper and lower parts. Graphene is in grey, the flake is colored according to the vertical displacement (white and blue protruding, red intruding). The coloring shows the formation of deformation waves from the tip contact point, propagating to the flake and triggering the horizontal motion.

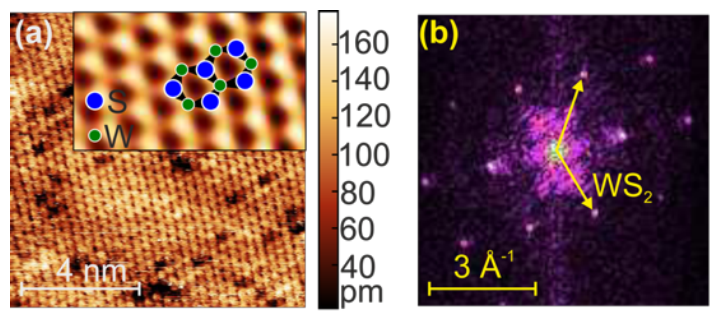

**Figure 5. Atomically resolved WS$_2$ on graphene.** (a) STM image (0.1 V tip voltage, 0.25 nA tunnel current) acquired on a WS$_2$ flake. Inset: zoom-in to the lattice (2D FFT filtered), with atom position overlay. (b) FFT of the raw image in reported (a).

# Supplementary Info

# Superlubricity of epitaxial monolayer $WS_2$ on graphene

Holger Büch[1], Stiven Forti[1], Antonio Rossi[1,2], Domenica Convertino[1,2], Valentina Tozzini[2] and Camilla Coletti[1,3]

1. Center for Nanotechnology Innovation @NEST, Istituto Italiano di Tecnologia, Piazza S. Silvestro 12, 56127 Pisa, Italy

2. NEST, Istituto Nanoscienze – CNR and Scuola Normale Superiore, Piazza San Silvestro 12, 56127 Pisa, Italy

3. Graphene Labs, Istituto Italiano di Tecnologia, Via Morego 30, 16163 Genova, Italy

**Large area STM scanning**

Large-scale STM scan displaying multiple $WS_2$ flakes. Typically, during consecutive scanning frames, a number of flakes start sliding over the graphene substrate. Here we report subsequent large area STM scans where only a few flakes slide, as this makes easier to follow the path of each single crystal. Panels (a)-(c) are exemplificative of sliding paths observed in consecutive scans. Flakes appears to move with similar probability uphill and downhill the atomic steps. In the top portion of the micrographs we highlight the motion of two flakes, circled in light blue and green. The flake within the light blue circle is initially found at the edge of a terrace, while in each subsequent scan is found stepping down one terrace. The flake within the green circle is, at the opposite, climbing terraces. In general, we observe that climbing of a terrace is more likely to occur for lower step heights (see panel (d)). In all cases, typical locking position are the edge of steps. In general, bigger clusters of $WS_2$ flakes appear to be steady.

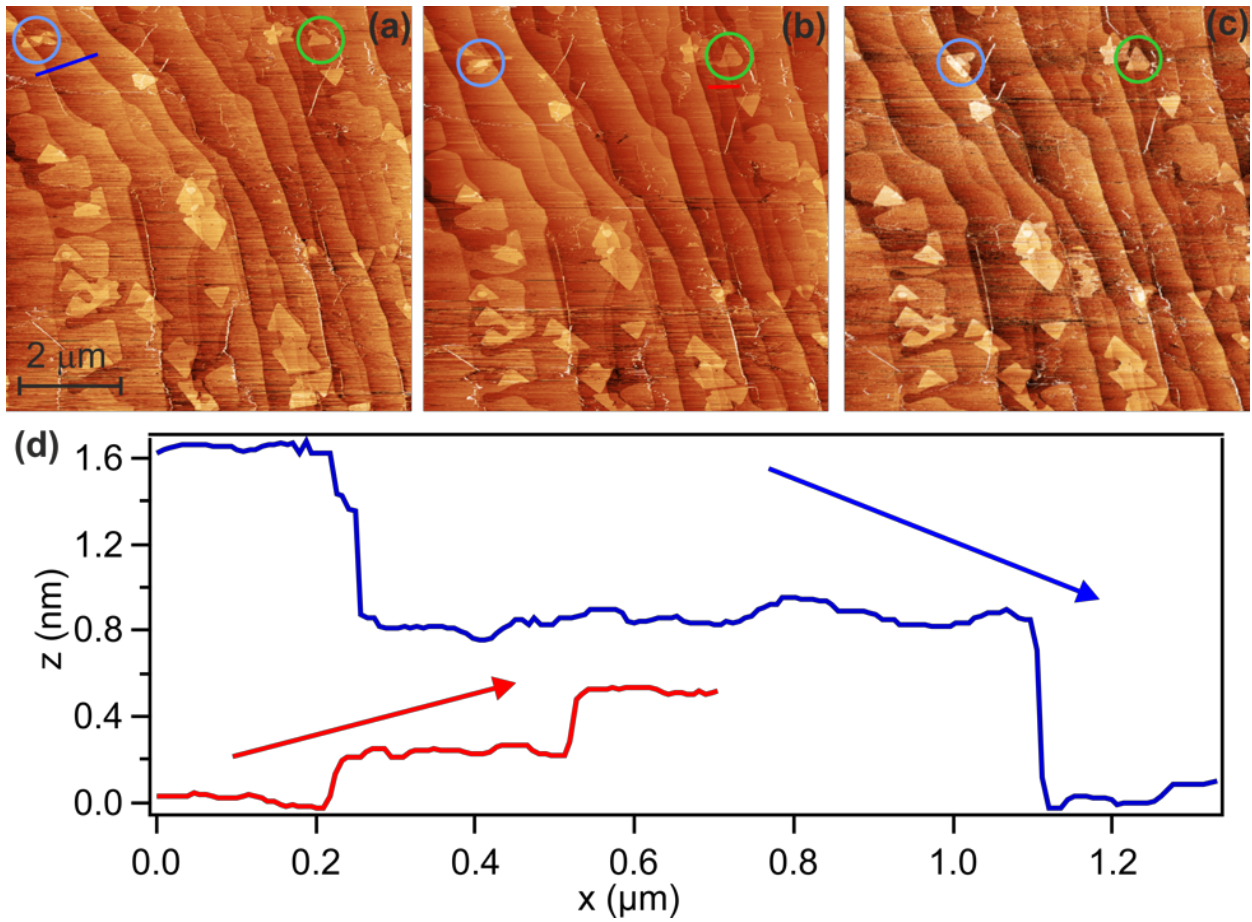

**Fig. S1.** Large area STM scan, showing WS$_2$ flakes moving in three consecutive scans (a) (b) and (c). Moving flakes are highlighted with colored circles. Images are obtained with tunnel current I=400pA, and bias voltage V=2.9V. Panel (d) shows the line profile of the terraces indicated in panel (a) and (b). It appears that the terraces climbed up have a lower height than those climbed down.

**AFM scanning**

AFM scanning in contact mode reveals the sliding of WS$_2$ flakes on graphene. Also when scanning with AFM, flakes are found to slide both in the downhill and in the uphill direction of the steps, although the former sliding direction appears to be more likely (see Supplementary video S7). Being the friction generated by the corrugation amplitude, in the first approximation the direction of motion should not make difference if the steps had an aspect ratio near to 1. In this case however the terraces are flat and large, which creates an asymmetry with respect to the direction of motion, favoring the downhill direction. Flakes often rotate around one point. Tip dragging effect can be excluded since the flakes are not moving accordingly with the scanning direction (see Supplementary video S6). Also, the sliding occurs simultaneously for different flake in the same frame. Moving flakes are often imaged with deformed darker contrast.

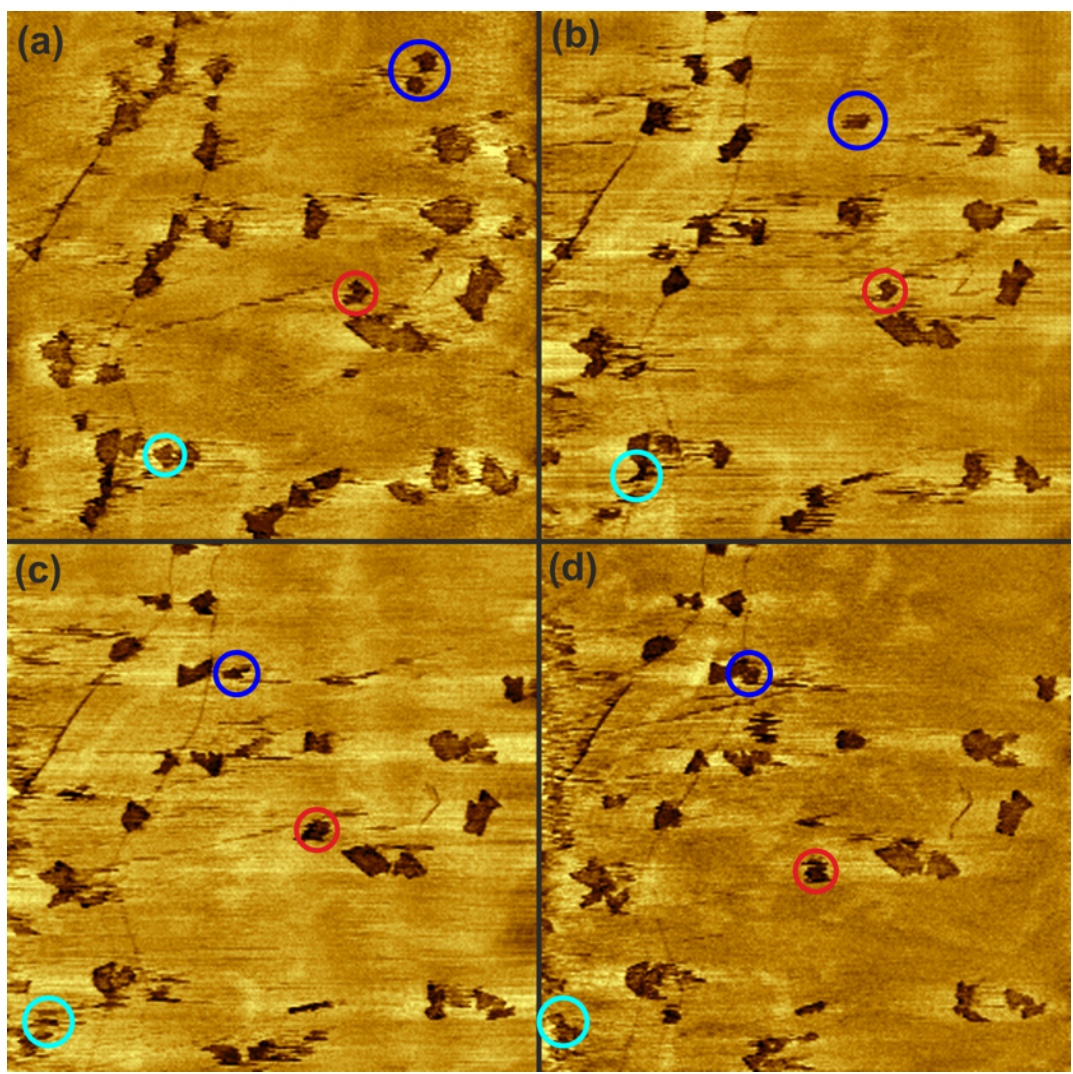

**Fig. S2.** AFM deflection scan in contact mode showing small flakes highlighted by colored circles moving in four consecutive scans (a-d). "Downhill" direction of terraces is from right to left. Image size is 10μm x 10 μm.

**Details of simulation and modeling**

The model systems used in this work are reported in Figure S3. Flakes of three different sizes from ~5 to ~20 nm are used, in combination with simulation sufficiently large supercells from 16 to 31 nm) including the substrate. The largest model system includes ~40500 atoms.

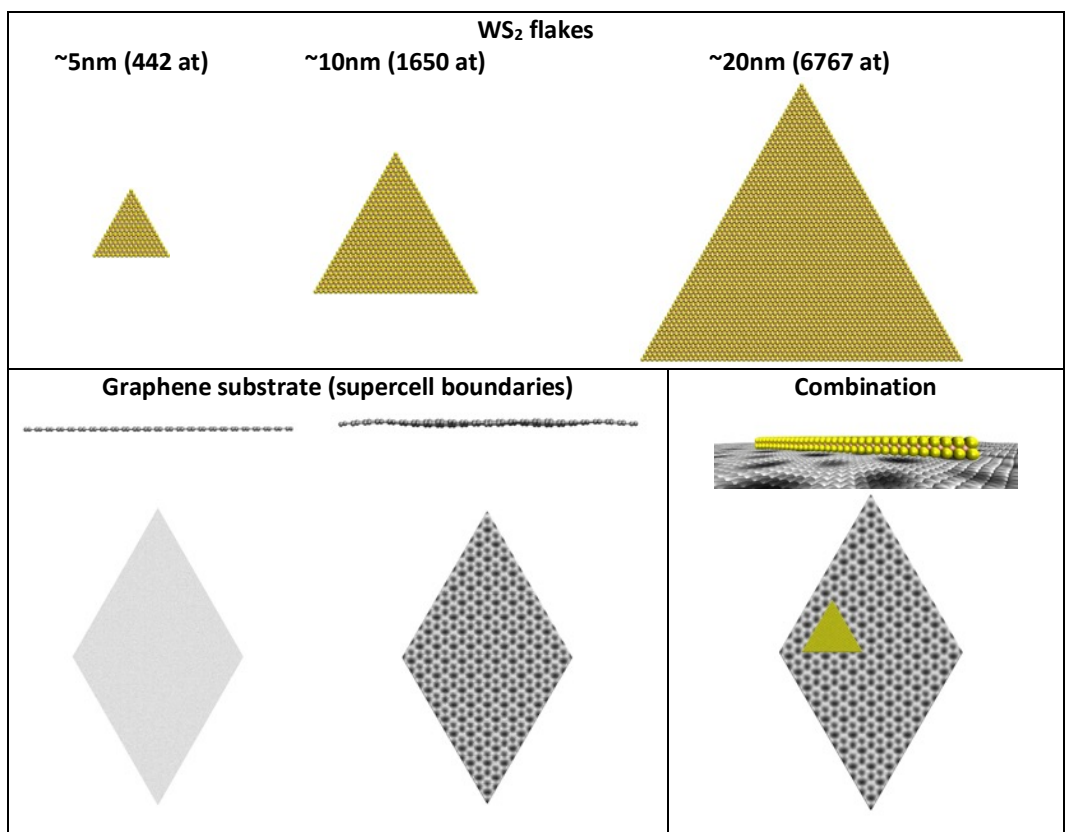

**Fig. S3.** Representation of the substrate (cut within the simulation supercell) and of the flakes of $WS_2$ (flakes are in scale between each other but not with the substrate structures). Substrates are considered both flat and corrugated with corrugation amplitude (bottom-to-top vertical displacement) of ~0.57Å at T=0K. Corrugations are highlighted with gray shadows (bright areas protrude, dark areas intrude).

In order to speed up simulations, we represented the intra-molecular interactions of the flakes and the substrate with elastic network models, defined by two parameters: the elastic constant and the number of neighbors included in the network. For the flakes, up to the third neighbors were included. The parameters were adapted from a similar model reproducing phonons in 2D $WS_2$ systems[i], using however bond angles and dihedral in place of second and third neighbor distances and therefore requiring a conversion of elastic constants. Therefore, elastic constant turn out to be distance dependent, as in "distance dependent elastic networks" widely used for biopolymers[ii]. The elastic constant of graphene, conversely, was fitted on fluctuation data derived by simulation with the Broido potential[iii], known to reproduce phonons in graphene[iv].

Graphene required a six-neighbor model, being single layer, as opposed to the three layers WS$_2$ flake.

In order to emulate the SiC substrate effect, the graphene sheet atoms were anchored to their starting location with a tethering potential also maintaining the corrugation typical of graphene on SiC. The pattern of corrugation was build following the one obtained in our previous DFT calculations[v], reproducing the experimental one. The maximum amplitude and the value of the elastic constant were adjusted in order to reproduce the experimentally measured values of the vertical displacement measured at high temperatures. This simplified, few parameters representation of the substrate, therefore, has the advantage of being capable of reproducing the experimental structural features of the graphene sheet on SiC (corrugation and dynamical out of plane fluctuations) with a limited computational cost. Graphene-flakes interactions are represented as Lennard-Jones with standard parameterization. Table S.1 reports a summary of the used parameters for simulations.

**Tab S.1.** Summary of the model parameters.

| Graphene sheet | | |
|---|---|---|
| Tethering potential | Harmonic, anchoring to the starting position | $k_t$=0.065 eV/Å$^2$ |
| Corrugation amplitude | | 0.57Å |
| Elastic Network | 6$^{th}$ neighbor ($r_{cut}$ = 5.2 Å) | k=34.7 eV/Å$^2$ |
| **WS$_2$ flakes** | | |
| Distance dependent Elastic Network: K=200 exp(-2.0*($r_0$-2.4)) | | |
| | $r_0$ (Å) | K (eV/Å$^2$) |
| First neighbor (S-W) | 2.40 | 8.67 |
| Second neigh (W-W) | 3.160 | 1.95 |
| Second neighbor (S-S) | 4.437 | 0.165 |
| | 3.164 | 1.91 |
| Third neighbor (S-W) | 3.972 | 0.373 |
| | 5.076 | 0.0433 |
| **Flake - graphene interaction** | | |
| Lennard Jones | E (meV) | σ (Å) |
| W-C | 2.91 | 3.14 |
| S-C | 7.50 | 3.56 |

Different sets of simulations were performed: (i) microcanonical ensemble dynamics starting from a locally optimized structure, at different average temperatures, (ii) vertical pulling of the flake to emulate the tip effect and subsequent free dynamics and (iii) overdamped dynamics in the presence of a uniform force applied to the flake in y (harmchair) direction. A summary of the performed simulations is reported in Table S.2 and the main results are described below for each set.

*Description of runs of type (i).*

In general, depending on the temperature and on the size, we can recognize three different classes of motion: a. oscillations about starting position; b. percolation, meaning a motion with frequent changes of direction and orientation, c. roto-translation motion, with a defined starting velocity of the CM (reported in the table). In addition the times of activation of percolation $\tau_p$ and of roto-translation $\tau_{rt}$, and the starting velocity upon activation of roto-translation, also depend on size and temperature. Therefore there is some arbitrariness in the definition of the activation temperature, depending whether the percolation is included or not and which activation time is considered. However, we can safely say that the 5 nm flake motion is already activated between 126 and 158 K (the latter reported in Figure S4, c,d), reporting the activation dynamics of the ~5 nm flake to completion of Figure 3 in the main text. At 104 K the 5 nm flake shows an extended percolation phase, in which the flake undergoes roto-translational hopping between different configuration, showing marked preference for orientation of 0 and 30° with respect to zig-zag edge of graphene. This run is reported in Figure S4 a,b. Conversely the 10nm flake is activated over at ~211. The 20nm flake run is within the percolation phase in a run length of 15ns at 295K, which is the predicted transition temperature. Interestingly enough, in the run at ~300K performed with the frozen substrate, the flake shows only an oscillatory motion, while at a similar temperature the percolation is activated almost instantaneously in the run with mobile substrate, immediately followed by roto-translation, indicating that the substrate dynamics has a fundamental role in the activation of motion. Movies of the runs of 5nm flakes at 104K and at 157 K are also reported as additional supplementary information. The first one shows that the flake has two preferential orientations, namely 0 and 30 deg, during the percolation phase. The second shows the starting of the roto-translational superlubric motion after the oscillation-percolation phase.

**Tab S.2.** Summary of performed simulations. $\varepsilon_{vdw}$ is the average van der Waals energy per contact S atom during the simulation (negative sign with respect to the completely detached flake configuration). $v_{CM}$ is the velocity of the center of mass at the beginning of roto-translational motion. $\tau_p$ and $\tau_{rt}$ are the times at which the percolation and roto-translational motions start, respectively. Micro-canonical simulations (labeled "µ") are started from a locally optimized configuration, setting a non-zero starting temperature by means of a random field of atomic velocities. Dragging simulations are performed applying a constant force in y direction (harmchair of graphene) to each atom of the flake, so that the total force is proportional to the surface of the flake. In addition, at each step the kinetic energy is zeroed, so that the motion is "adiabatic" and the energy profile is mapped. This over-damping results in a constant sliding velocity. The last simulation is composed of two parts: In the first, a vertical (along z) pulling force (with over-damped dynamics) is applied to a few atoms in a decentered location of the flake, not to generate geometric bias; once the deformation reached the value of ~2.5 Å, the force was turned off and the system let free to evolve in microcanonical ensemble, with the starting velocity set at RT as in the other µ canonical simulations. The starred runs are those used for Figures 3 and 4 in the main text. Double starred run are reported in Fig S.4. Movies of the dotted runs are reported as supplementary information (see below for the description).

| | Flake Size (nm) | Simulation type | T (K) | Run length (ps) | $\varepsilon_{vdw}$ (meV) | $v_{CM}$ Å/ps | $\tau_p$ (ns), $\tau_{rt}$ (ns) | Motion description |
|---|---|---|---|---|---|---|---|---|
| (i) | 5.05 | µ, frozen sub | 301 | 100 | -108 | 0 | - | Oscillations |
| | 5.05 | µ | 52 | 20000 | -111 | 0 | - | Oscillations |
| | 5.05 | µ ● | 104 | 30000 | -110 | 0 | ~2 | Oscillations, percolation |
| | 5.05 | µ ** | 126 | 10000 | -110 | 0.015 | ~1, ~5 | Oscillations, percolation, roto-translations |
| | 5.05 | µ ** ● | 158 | 10000 | -109 | 0.07 | <0.001, ~2-6 | Percolation followed by roto-translation |
| | 5.05 | µ | 319 | 2000 | -107 | 0.1 | <0.001, ~1 | Percolation followed by roto-translation |
| | 10.1 | µ | 157 | 10000 | -110 | 0 | - | Oscillations |
| | 10.1 | µ * | 211 | 5000 | -109 | 0.09 | <0.001, ~3 | Percolation followed by roto-translation |
| | 10.1 | µ | 320 | 1000 | -108 | 0.2 | <0.001 | Roto-translation |
| | 20.2 | µ | 295 | 1500 | -109 | - | ~1 | Percolation |
| (ii) | 20.2 | Pulling + µ * | 325 | 10000 | -108 | 0.07 | ~0.2 | The flake relaxes for 200 ps the starts before activation of roto-translational motion |
| (iii) | 5.05 | Dragging * | - | 100 | -113 | 0.4 | - | Constant sliding velocity |
| | 10.1 | Dragging * | - | 100 | -111 | 0.4 | - | Constant sliding velocity |

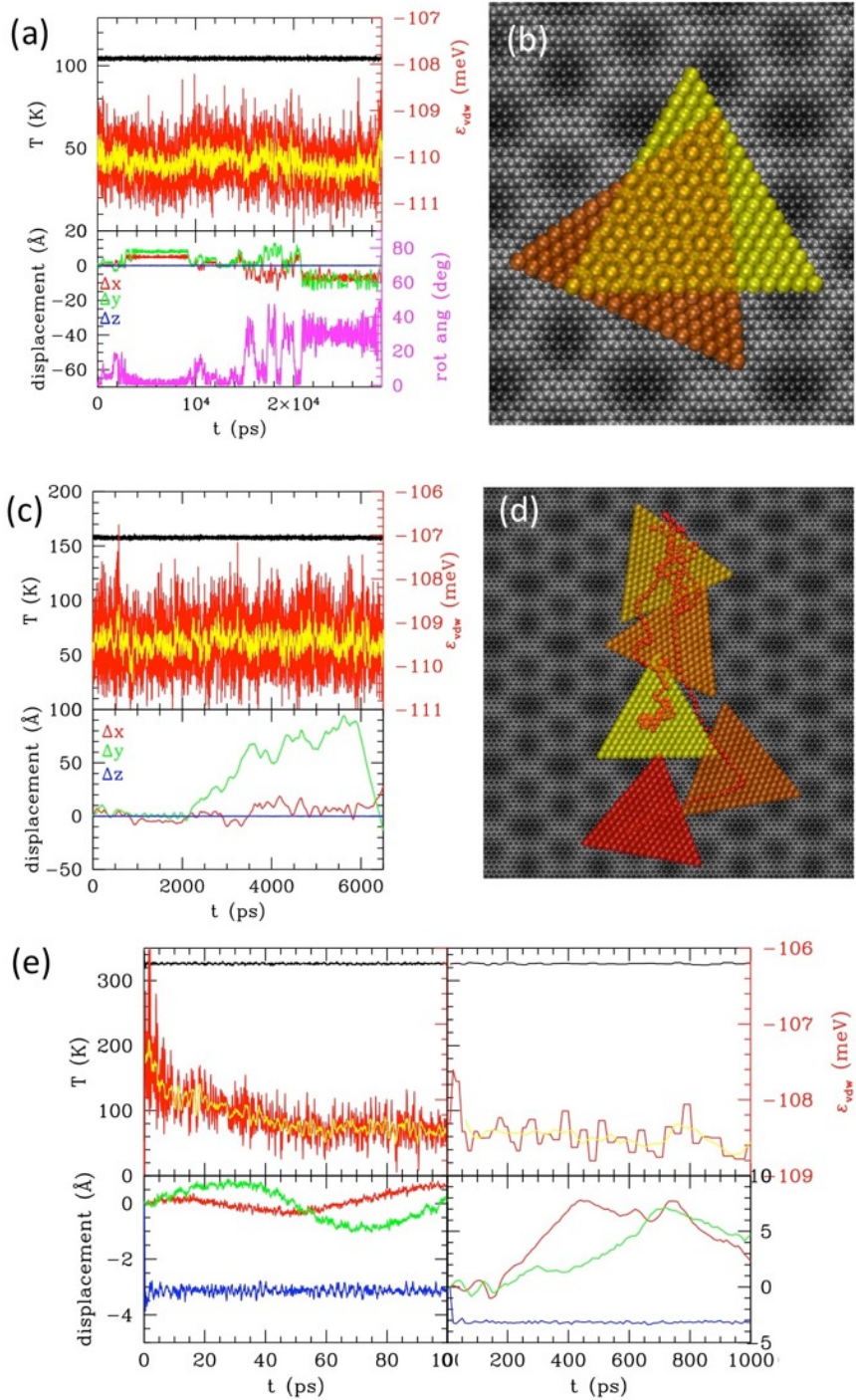

**Fig. S4** Description of selected runs of Table S.2. (a) reports the run of the 5nm flake at 104K. The temperature (black line) and vdW energy profiles (red, raw data, yellow, running average) per contact atom; in red, green and blue in the bottom plot the three Cartesian component of the displacement of the center of the flake are reported. The rotation angle is reported in magenta in the bottom plot (scale on the right vertical axis). Percolation is detectable as jumps in the x and y coordinates, while the rotation angle is clearly seen to jump between two orientations, 0 and 30 deg. These are also shown in yellow and red in (b) where the system is represented vdW spheres representation (the substrate colored according to the height, light grey = protruding). (c) Illustrates, for the run of the flake of ~5nm at ~157K, with the same color-coding of lines as (a) The oscillation/percolation phase is visible up to 2 ns, lasting for the x coordinate at 6ns. The ballistic-like motion starts afterwards. The trajectory with some snapshots is reported in (d). (e) illustrates the run of the 20nm flake following the pulling. Color notation is the same as (a) except that the displacement of the pulled atom is reported instead of that of the center of the flake. During the early phase (left part, first 100ps) the immediate relaxation of z coordinate is visible and related oscillations of x and y coordinates; during the first 100 ps also the vdW component of the energy relaxes; the right part of the plot shows the first ns of simulation (smaller data writing frequency). The activation of horizontal

motion occurs at ~200ps after a couple of oscillations. The trajectory and snapshots from this run are reported in Fig 4 of the main text.

*Description of runs of type (ii).*

The run of type (ii) is described in the main text. Figure S4 (e) reports energy profiles and structural parameters along the simulation. As said, the simulation is separated in two parts, the first is the emulation of the tip pulling, followed by free dynamics.

*Description of runs of type (iii).*

The two runs of type (iii) are already described in the main text, Figure 3c.

**Supplementary movies**

*S5_104K.mov*

Reporting the first 1.8ns of the first dotted simulation of Table S.2

*S5_158K.mov*

Reporting the first 6ns of the second dotted simulation of Table S.2

*S6_SupplementaryAFMVideo.mp4*

Video built with subsequent AFM deflection scans. Each image was obtained with a trace and retrace movement of the tip. Trace and retrace images are identical. In this video flakes slide along different directions. Lateral size of video is 5 x 5 μm.

*S7_SupplementaryAFMVideo.mov*

Video built with subsequent AFM height scans. Each image was obtained with a trace and retrace movement of the tip. Trace and retrace images are identical. In the central part of the image (slightly right-hand side, just above a cluster of flakes) a WS2 flake is seen moving uphill a terrace (about 0.8 nm heigh, see Fig. S8). Lateral size of video is 10 x 10 μm.

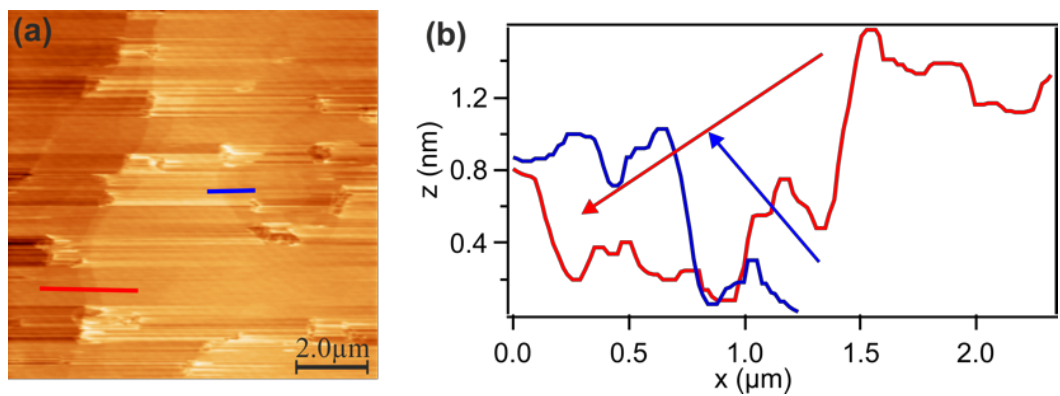

**Fig. S8.** AFM image from video S7. The blue line profile indicates the height of the terrace which is climbed up by the flake positioned just below (on the hand-right-side) the blue line. The red line profile indicates the height of the step climbed down.